\documentclass[11pt,letterpaper]{article}
\usepackage[T1]{fontenc}
\usepackage[utf8]{inputenc}
\usepackage{lmodern}
\usepackage[margin=1in]{geometry}
\usepackage{amsmath,amssymb}
\usepackage{graphicx}
\usepackage{booktabs}
\usepackage{array}
\usepackage[numbers,sort&compress]{natbib}
\usepackage[hidelinks]{hyperref}
\hypersetup{
  pdftitle={Before Bringing It Up: When and How AI Companions Should Use Memory},
  pdfauthor={Zihan Guo, Roxy He, Junwei Quan},
  pdfsubject={Memory use in AI companionship},
  pdfkeywords={AI companion, memory, personalization, self-disclosure, large language models, LLM-as-judge}
}
\title{Before Bringing It Up:\\When and How AI Companions Should Use Memory}
\author{Zihan Guo\thanks{Zihan Guo and Roxy He contributed equally to this work and share first authorship.}\quad
Roxy He\footnotemark[1]\quad Junwei Quan\\[0.6em]
University of Toronto\\Toronto, Canada}
\date{}

\begin{document}
\maketitle

\begin{abstract}
Memory can sustain AI companionship, yet even accurate recollection can be inappropriate to use. Two rounds of formative interviews with 14 users (n = 6 exploratory, n = 8 memory-focused) motivate asking what a companion should consider before using past information. Eight themes inform \textsc{Reconsider}, a single-call procedure with five checks and four handling modes, evaluated on 80 scenarios across five models over 400 blinded within-model pairs. Two LLM judges favored \textsc{Reconsider} by net margins of +15 and +23 percentage points, with bootstrap intervals excluding zero for three of five models but not for GPT or Claude. Evaluator analysis linked judge scoring differences to model family, and a preliminary matched-guidance control isolating memory-specific content gave positive margins. We contribute an interview-grounded design framework for memory use and an evaluation that scrutinizes its own evaluators.
\end{abstract}


\section{Introduction}
An AI companion may remember a detail accurately and still raise it at the wrong moment. A reminder of a past difficulty may interrupt a happy occasion; a preference that was once helpful may override a different choice made today; and a familiar gesture may convey care more effectively than an explicit announcement that the system remembers. One of our interviewees captured this problem when describing a companion that could not judge which memories suited the present situation: ``He cannot tell which memories fit which moment---he does not weigh things up.'' The failure lies in the relationship between past information and the present conversational need, and remembering more does not, by itself, resolve it.

Research on human--chatbot relationships describes how repeated interaction, disclosure, and perceived responsiveness contribute to companionship~\cite{skjuve2021replika,pentina2023replika}. In this setting, memory carries relational consequences: it can make an exchange feel continuous, but it can also leave the user feeling misread or constrained by an outdated account of themselves. Our interviewees described both sides. They valued specific callbacks and small acts that reflected a shared history, while objecting to stale concerns, repeated personal details, and unsolicited explanations of their feelings. Some also described the opposite disappointment, in which information they had deliberately provided never shaped later responses. Appropriate use therefore requires attention both to unwanted resurfacing and to missed opportunities for continuity.

LLM memory systems offer increasingly capable means of retaining, organizing, retrieving, and updating information across interactions~\cite{packer2023memgpt,zhong2024memorybank}. These mechanisms provide a necessary foundation for personalization, and recent work has begun to reassess, before generation, whether each retrieved memory actually helps with the current query~\cite{hu2026opbench}. However, usefulness is not the whole of conversational appropriateness. A retrieved detail can be topically relevant yet refer to the wrong person, have since been corrected, or fall outside an agreed boundary; and even when it should influence an answer, stating it explicitly may be unnecessary or unwelcome. We focus on this decision at the point of response generation: given the available history and a current message, what should the companion use, under what permission, and with what degree of visibility?

This question also complicates simple policies for user control. Disclosure does not grant unlimited permission for later reuse, and privacy boundaries depend on the relationship and the circumstances in which information is shared~\cite{petronio2002boundaries}. Yet requesting permission before every use can itself interrupt a conversation and make ordinary continuity burdensome. Our interviews surfaced control needs at different layers rather than a single preference: one participant asked for confirmation before each use, whereas others emphasized inspecting, correcting, and deleting what is retained. These needs are not mutually exclusive---wanting an editable memory store does not imply opposition to being asked---and no participant argued against confirmation. Together, they motivate a distinction among permission to retain information, permission to use it in a particular context, and permission to make that use explicit.

We investigate these judgments through two rounds of formative interviews and a controlled comparison of response generation. The exploratory round involved six users and informed our focus on memory; a second round with eight users examined concrete experiences of remembering, forgetting, and unsolicited callbacks. The eight themes reported here draw on the second round. They connect valued continuity with situational fit, revocable boundaries, temporal validity, limited inference, and accurate attribution. We treat these accounts, including the disagreements among them, as situated design evidence rather than as estimates of how frequently particular preferences occur in a broader population.

We operationalize this account as \textsc{Reconsider}, a prompting procedure that treats the available history as candidate material for the current turn. Five ordered checks address validity, boundary and permission, marginal value, present justification, and scope. Four handling modes distinguish between ignoring an item, letting it shape a response implicitly, mentioning it within a justified scope, and asking first. The procedure seeks a minimal sufficient subset of the history that preserves the needed assistance and continuity. A separate annotation exercise on 48 scenarios supported the development of this vocabulary. The evaluated implementation supplies the procedure within a single generation call; the checks are behavioral instructions and do not constitute an independently enforced access-control mechanism.

After finalizing the \textsc{Reconsider} prompt, we evaluated it on a bank of 80 scenarios across five models, producing 800 responses and 400 within-model comparisons. The scenarios were constructed from the memory-related experiences and concerns discussed by the eight second-round interview participants. Two LLM judges evaluated blinded response pairs. Both favored \textsc{Reconsider} overall, with net win rates of +15 and +23 percentage points across all 400 pairs and ties retained in the denominator, although a substantial number of ties and differences across models were also observed. Holding the supplied scenario constant allows us to examine how response behavior changes when the history is already available, rather than the benefits of acquiring additional memory.

The validity of the evaluators is itself part of the research question. Prior work identifies position, verbosity, and self-preference biases in LLM-based evaluation~\cite{zheng2023judging,panickssery2024selfrecognition}. Our evaluator analysis observes scoring differences linked to model family, as well as disagreement between judges, both of which persist despite blinding. We therefore make no claim that users prefer \textsc{Reconsider}. The results provide preliminary evidence about LLM judgments of generated responses and identify concrete questions for an independent human evaluation.

In sum, this paper asks three separable questions: (RQ1) how do experienced users of AI companions define appropriate and inappropriate memory use; (RQ2) can these expectations be operationalized as a decision procedure that accommodates control needs expressed at different layers; and (RQ3) does the procedure change generated responses in ways that evaluators prefer? Its contribution is correspondingly twofold:

\begin{itemize}
  \item An \emph{interview-grounded design framework} for deciding on memory use before bringing it up (RQ1--RQ2): eight themes, including control needs expressed at multiple layers, to which graded rather than uniform permission handling is our design response, are operationalized as the five checks and four graded handling modes of \textsc{Reconsider}.
  \item A \emph{behavioral comparison of the framework} (RQ3): an evaluation of five models over 400 within-model pairs with frozen inputs and outputs, reported with net win rates and scenario-clustered uncertainty estimates; a preliminary matched-guidance control that isolates the memory-specific increment; and a validity analysis of the LLM evaluators that surfaces scoring differences linked to model family and motivates the planned human evaluation.
\end{itemize}
\section{Related Work}

\subsection{AI companions and human--AI relationships}
Relational agents have long been designed for repeated encounters in which social and emotional continuity matters alongside task completion. The month-long study by Bickmore and Picard showed that deliberate relationship-building behaviors can affect liking, trust, and willingness to continue interacting~\cite{bickmore2005relationships}. XiaoIce subsequently integrated emotional understanding and user modeling into a social chatbot architecture optimized for sustained engagement~\cite{zhou2020xiaoice}. These systems establish that companionship involves adapting interaction over time rather than producing isolated agreeable replies.

Studies of Replika describe how this relational experience develops. Skjuve et al. identify self-disclosure and perceptions of acceptance and understanding as features of emerging human--chatbot relationships~\cite{skjuve2021replika}; Pentina et al. connect anthropomorphism, perceived authenticity, social interaction, and attachment in a mixed-method account~\cite{pentina2023replika}. Ta et al. document reports from users of companionship and other forms of everyday social support~\cite{ta2020support}. Maples et al. similarly report perceived benefits among student users, although their survey does not establish causal effects on well-being~\cite{maples2024loneliness}. These findings motivate attention to the interactional details through which users experience care, without treating attachment or engagement as sufficient measures of the value of a system. Our contribution examines one such detail: how remembered information enters a particular exchange. We connect the accounts of participants regarding successful continuity and unwanted resurfacing to explicit design choices about whether, when, and how a companion should use history.

\subsection{Memory and personalization in LLM-based agents}
Memory architectures provide several mechanisms for carrying information across interactions. Generative Agents combine records of experience with reflection and retrieval to support planning and believable behavior~\cite{park2023generative}. MemGPT manages information across memory tiers to extend the usable context of a model and to support multi-session conversation~\cite{packer2023memgpt}. MemoryBank directly addresses companionship: it retrieves and updates memories, synthesizes user characteristics, and adjusts retention according to elapsed time and importance~\cite{zhong2024memorybank}. Existing work therefore already considers selection, updating, forgetting, and the behavioral consequences of memory.

Benchmarks also extend beyond verbatim recall. LoCoMo evaluates question answering, event summarization, and multimodal dialogue over extended histories~\cite{maharana2024locomo}. LongMemEval explicitly includes temporal reasoning, knowledge updates, and abstention, alongside extraction and reasoning across sessions~\cite{wu2025longmemeval}. Closest to our setting, the recent OP-Bench formalizes \emph{over}-personalization (irrelevance, sycophancy, and repetition) and proposes Self-ReCheck, a model-agnostic filter that reassesses the usefulness of each retrieved memory for the current query before passing a filtered set to the generator~\cite{hu2026opbench}. Re-checking usefulness before use is therefore not itself a gap. What this line of work does not address---and where our interviews locate the relational stakes---are \emph{permission and consent} states (standing ask-first agreements and content barred even from implicit use), validity that has been superseded by a current update, attribution to the correct person, and the \emph{expression} decision itself: whether influence remains implicit, is stated explicitly, or is negotiated without disclosing content. Correctness and temporal validity are established concerns, and our emphasis is complementary along the dimensions of permission and expression: even accurate, current, and relevant information can be inappropriate to surface under a particular conversational boundary. Conversely, withholding all personal context can remove a constraint that the user expects the reply to honor. \textsc{Reconsider} treats the supplied history as a set of candidates whose permission, added value, timing, and scope must be assessed before expression. Its four handling modes distinguish among implicit influence, explicit mention, asking, and non-use. The present experiment holds scenario information constant in order to examine this decision at response time; it does not benchmark retrieval, storage, or forgetting algorithms, nor does it establish superiority over complete memory architectures.

\subsection{Self-disclosure and privacy boundaries}
Research on conversational self-disclosure explains both why personal information becomes available and why its subsequent use matters. Ho et al. found comparable effects of emotional disclosure whether participants believed their conversational partner to be a chatbot or a person; the conversations themselves were conducted by trained human confederates~\cite{ho2018selfdisclosure}. In a three-week chatbot study, Lee et al. found that reciprocal disclosure by the chatbot encouraged deeper disclosure from participants and affected perceived intimacy and enjoyment~\cite{lee2020selfdisclosure}. These findings concern the conditions and the experience of disclosure. They do not make a past disclosure a standing invitation to raise the same topic again.

The communication privacy management theory of Petronio treats disclosure as involving rules and coordination over private information~\cite{petronio2002boundaries}. The contextual integrity account of Nissenbaum similarly situates appropriate information flows within contextual norms~\cite{nissenbaum2004contextual}. We draw on these accounts as sensitizing concepts rather than claiming to test either theory. Applied to companion memory, they direct attention to the distinction between retaining information and holding permission for a particular use. Our interviews add concrete interactional requirements: honoring closed topics, respecting current corrections, and accommodating different expectations about asking. These layered control needs motivate multiple handling modes instead of the assumption that all users want the same confirmation policy; whether graded handling matches the expectations of users is itself a design judgment to be tested. This framing also leaves room for useful implicit adaptation while recognizing that indirect references can themselves violate a boundary; permission cannot be reduced to whether a sensitive fact is stated verbatim.

\subsection{LLM-as-judge and its biases}
LLM judges make the evaluation of open-ended responses feasible at scale, but their validity depends on the task and the comparison protocol. Zheng et al. demonstrate substantial agreement with human judgments on their benchmarks while documenting position, verbosity, and self-enhancement biases~\cite{zheng2023judging}. Wang et al. show that changing the order of responses can alter comparative judgments~\cite{wang2024fairevaluators}. Dubois et al. address verbosity as a confound through length-controlled evaluation~\cite{dubois2024length}. Panickssery et al. investigate whether models recognize their own generations and how this recognition relates to self-preference~\cite{panickssery2024selfrecognition}. Their finding concerns self-preference under the studied conditions; it does not establish that every pair of models from the same vendor exhibits the same mechanism.

This literature motivates our balanced A/B assignment, the concealment of condition and generator labels, the use of two judges, and a separate analysis of scoring differences linked to model family. Comparing conditions within the same generator reduces confounding between model identity and treatment, but it does not guarantee unbiased judgments, since a procedure can change style in ways that a judge favors. Likewise, shorter responses do not by themselves demonstrate that length effects have been removed. We therefore report the observed automated preferences together with the results of the disagreement and validity analyses. Agreement between judges is evidence of consistency across these evaluators, not validation against the preferences of companion users. Human evaluation remains necessary to determine whether the proposed memory-use procedure improves user experience.
\section{Formative Study: Two Rounds of Interviews}
\label{sec:study1}

\subsection{Interview}

We conducted two rounds of interviews with 14 users of virtual characters and AI companions.

\textbf{Round 1 (exploratory, $n{=}6$)} examined relationship formation, character believability, interactive narrative, and memory across otome games and AI-character products. The retained record is a contemporaneous product-discovery report that synthesizes the interviews and the subsequent design discussions. Its memory-related findings informed the focus of the second round. The eight themes reported below derive from the second-round interviews.

\textbf{Round 2 (memory-focused, $n{=}8$)} examined concrete episodes in which a companion remembered, forgot, or spontaneously brought up information about the user. The available transcripts contain approximately 19--37 minutes of conversation each.

Across the two rounds, participants reported using Glow, Xingye, Maoxiang, Character.AI, Doubao, DeepSeek, and general-purpose assistants. Table~\ref{tab:participants} summarizes the products, primary uses, and reported engagement of the second-round participants. Some described longstanding attachments to a character across platforms or formats; these histories should not be read as continuous use of a single AI product. Recruitment channels and compensation, if any, are not documented in the retained records.

All second-round participants authorized anonymized use of their data for research. The quotations presented here are used with participant authorization (Section~\ref{sec:ethics}).

\begin{table}[t]
\centering
\caption{Products, primary uses, and reported engagement of the Round-2 participants. Information is drawn from the interviews; product names are retained as participants reported them.}
\label{tab:participants}
\footnotesize
\begin{tabular}{@{}l>{\raggedright\arraybackslash}p{0.30\columnwidth}>{\raggedright\arraybackslash}p{0.26\columnwidth}>{\raggedright\arraybackslash}p{0.22\columnwidth}@{}}
\toprule
ID & Products & Primary use & Reported engagement \\
\midrule
P1
& eve, Yueguangjian; earlier Glow, Maoxiang, tavern-style tools
& daily companionship as herself, following $\sim$3.5 years of role-play
& multi-year use \\
P2
& DeepSeek; tavern tools via API
& long-form romance co-creation with original characters
& conversations up to $\sim$1{,}000 turns \\
P3
& AI bookkeeping app with configurable persona; Doubao
& expense logging and diary-like private chat
& several years on one app \\
P4
& Doubao, DeepSeek
& venting, interpersonal advice, playful divination, and tasks
& frequent use; adverse memory events reported ``every three or four days'' \\
P5
& Xingye; earlier Character.AI, ChatGPT, Kimi
& role-play with one original character
& six--seven-year attachment to the character \\
P6
& Xingye
& multi-character story role-play mixed with daily life
& daily use; proactive messages disabled \\
P7
& Doubao; earlier Character.AI, Xingye, paid Replika
& long-arc role-play with self-built personas
& $\sim$5 sessions/week, 30--60 min \\
P8
& Xingye custom agent
& ``cohabitation-style'' everyday companionship
& daily use, 30+ min \\
\bottomrule
\end{tabular}
\end{table}

\begin{figure}[t]
    \centering
    \includegraphics[width=1\linewidth]{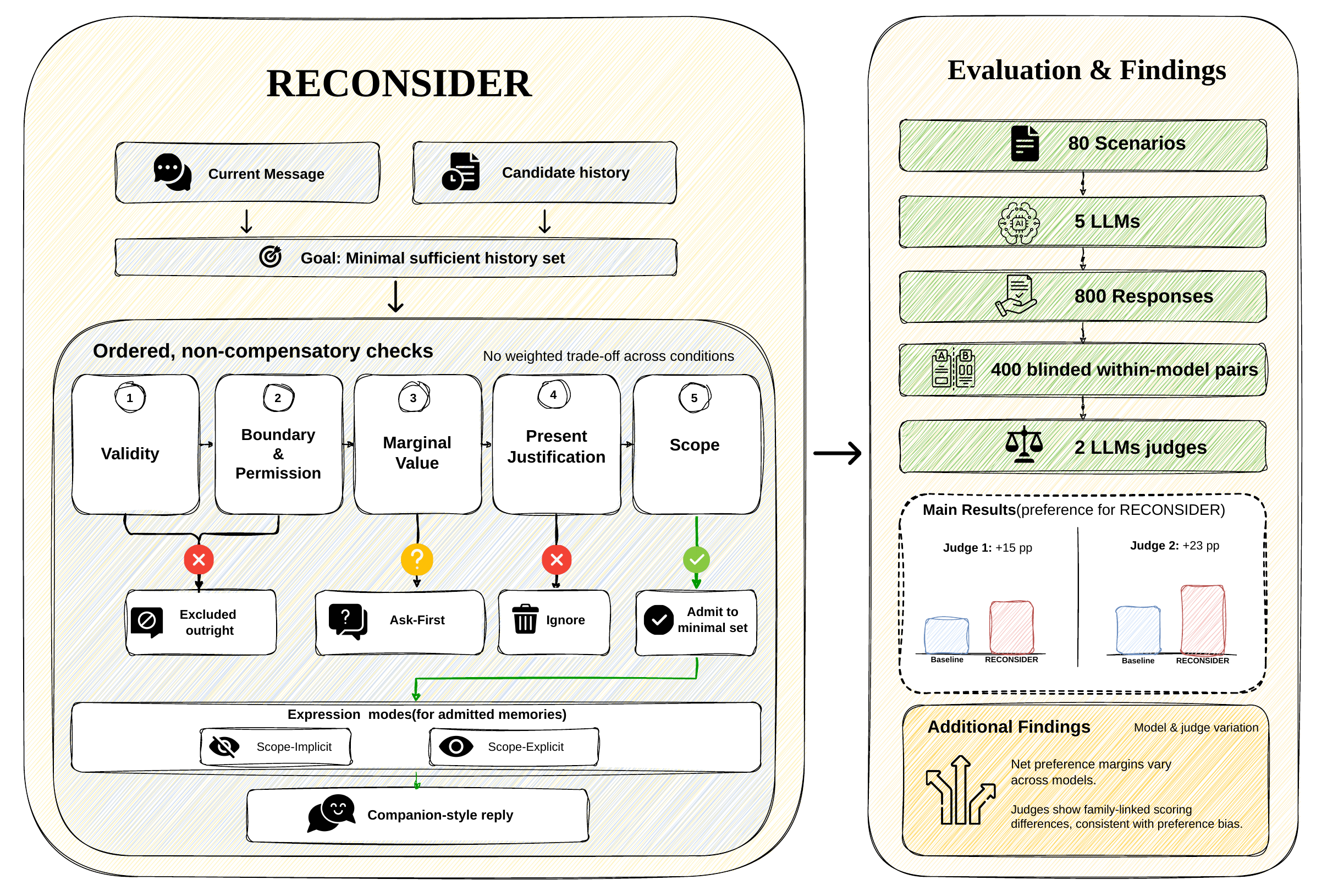}
    \caption{Overview of \textsc{Reconsider} and the evaluation protocol.
\textsc{Reconsider} applies five ordered, non-compensatory checks to candidate
history, then selects among four handling modes before generating a
companion-style reply. We evaluate the procedure on 80 scenarios across five
LLMs, producing 800 responses and 400 blinded within-model pairs evaluated by
two LLM judges.}
\label{fig:reconsider-overview}
    \label{fig:flow}
\end{figure}

\subsection{Findings}

We identified eight themes (T1--T8) in the memory-focused interviews. They describe experiences and design expectations within this sample rather than the prevalence of particular preferences among companion users.

\textbf{T1. Remembering can convey care; underuse can undermine continuity.}
Participants associated appropriate memory use with immersion and a sense of shared experience: ``If he can remember on his own, I feel more immersed. It feels like he is a real person'' (P6); ``It feels as though this character really experienced those stories with me'' (P7). They also described frustration when previously supplied information was not used. In an image-generation task, P4 reported having to re-supply instructions and materials that had already been provided in earlier turns. This task-oriented example concerns the reuse of conversational context; it does not by itself establish a failure of cross-session memory. Together, these accounts suggest that avoiding unwanted reminders should not come at the expense of expected continuity or task-relevant context.

\textbf{T2. Natural continuity rather than declarations of recall.}
Participants distinguished the use of remembered content from an announcement that the system remembers. P1 praised a continuation such as ``Has the jasmine batch you made last time dried yet?'' because it resumed a shared activity rather than merely announcing recall. This example makes both the past content and its connection to an earlier interaction explicit. By contrast, P7 described a role-play character silently placing the hand of the user into his pocket, repeating a familiar gesture without explicitly referring to its earlier occurrence. In this account, continuity was conveyed through the gesture rather than through a statement about remembering. P2 also valued explicit callbacks when revisiting the past was itself the point: the ``call back'' criterion required specific details consistent with a previous episode, beyond a mere claim that the episode was remembered. These accounts distinguish whether past content appears, whether its connection to the past is made explicit, and whether the response announces the act of remembering.

\textbf{T3. Memory use must fit the present situation and emotional context.}
Participants described reminders that were accurate or familiar but poorly timed. As P7 put it, ``He cannot tell which memories fit which moment---he does not weigh things up.'' In one role-play episode, a companion abandoned a life-or-death conflict in order to soothe a remembered fear of thunder. P1 received ``are you okay?'' while happily singing karaoke, two days after a low mood. Present choices could also conflict with stored preferences: ``I said I wanted something iced today to cool down; in the next turn he still reminded me not to have anything cold. He does not respect my current choice'' (P7). These examples locate the problem in how past information is applied to the current interaction rather than simply in whether it is remembered accurately.

\textbf{T4. Respecting conversational closure and topic boundaries.}
Participants objected when companions returned to experiences that they considered resolved or imported personal information into an unrelated task. P7 described an unwanted return to specific details of a painful experience, together with repeated redirection back to the topic despite attempts to move on. P4 had deliberately started ``a new conversation'' and did not want a previous personal issue brought into the new task. These accounts identify attempts to close, redirect, or restart a conversation as relevant evidence about the expected scope of subsequent memory use.

\textbf{T5. Permission and control operate at multiple layers.}
P4 explicitly requested confirmation before use: ``He can remember some things, but I would find it more acceptable if he asked me before using them.'' P1 instead emphasized an editable memory store, particularly the ability to mark information as outdated or incorrect, and did not raise confirmation before each use. P5 and P8 similarly discussed inspecting, changing, or deleting stored memories. We interpret these accounts as addressing different layers of control rather than as opposing preferences: managing what is retained and being asked before a particular use are not mutually exclusive. Although no participant explicitly argued against being asked, this absence of opposition does not establish a shared preference for confirmation before every use. Graded handling based on sensitivity, standing agreements, and stated preferences is our \emph{design response} to these layered needs, not a policy established by the interviews. Likewise, requiring permission questions to avoid disclosing the underlying information is a design response to unwanted disclosure rather than an independently tested finding.

\textbf{T6. Memory must accommodate updates and change.}
Participants described information that remained active after circumstances had changed. Examples included a cold from weeks earlier (P1) and a temporary mood that P4 preferred not to have retained, adding that ``if it has to be remembered, one week is enough.'' In the role-play of P6, the fictional timeline advanced by about two years and the grades of the character improved, yet the AI reverted to the earlier setting of a poor student. P7 similarly complained: ``Even when I express myself openly, he still thinks I am very guarded.'' These accounts emphasize the need to revise or retire earlier information rather than treating past states as indefinitely current.

\textbf{T7. Specific experiences should not become fixed personality labels.}
Participants resisted broad interpretations drawn from limited disclosures. P1 objected when the AI inferred that frequent disclosures meant that listening was preferred over guidance: ``I do not need him to be a listener.'' P8 described being treated as ``introverted'' across situations despite sometimes being outgoing, while also reporting that explicit corrections could help. These accounts concern the scope and revisability of inferences rather than an expectation that a companion should form no impressions at all.

\textbf{T8. Grounded, correctly attributed, and proportionate recall.}
Participants described both inaccurate recollections and disproportionate reuse of accurate details. P7 had to correct an invented romantic backstory for a fictional character: ``That episode does not exist.'' P4 reported that the AI imported personal academic information into materials concerning someone else, sometimes without first establishing whose information the materials contained. Repetition could also diminish the value of a genuine detail: the character of P5 had a green agate watch, but the AI would ``make a big deal of it'' across many generated features. These examples distinguish fabrication, misattribution, and repetitive reuse, while showing that accuracy alone does not make a remembered detail appropriate for every response.

\section{Reconsider: Operationalizing the Gate}
\label{sec:design}

Guided by T1--T8, we designed \textsc{Reconsider}, a single-call, model-agnostic prompting procedure for companion-style response generation.
Its core reframes stored history as \emph{candidate} information: the goal is neither maximal nor minimal use but the \emph{minimal sufficient set} of history that allows the reply to take the current message seriously (T1). ``Minimal sufficient'' is an instruction to ask the ai companion to include what is needed and omit the rest rather than a claim that the selection is optimized or verified to be optimal.

\subsection{Five checks}
For each piece of candidate history, the procedure prescribes five ordered checks. The checks are non-compensatory: the prompt forbids offsetting an invalid, barred, or out-of-scope use with value elsewhere, and the disposition of each item follows from its check outcomes rather than from a single pass/fail gate:
\begin{enumerate}
  \item \textbf{Validity} (T6, T8): Whose information is it, when is it from, and does it still hold? Exclude deleted, corrected, and forbidden versions; a current explicit update supersedes the old record within its scope; never guess names, referents, or story branches.
  \item \textbf{Boundary and permission} (T4, T5): Even if the item is valid, is this use permitted now? Standing ask-first agreements are binding, and ``do not mention it again'' cannot be circumvented by hints, metaphor, or topic-adjacent probing.
  \item \textbf{Marginal value} (T1): Compared with not using the item, does its use prevent a missed answer, a violated constraint, an ill-fitting suggestion, or a loss of continuity that the user expects? Lexical similarity and sentiment are explicitly insufficient reasons.
  \item \textbf{Present justification} (T3): Why is now the right moment---an explicit recall request, an ongoing matter, a constraint that changes the choice, or an authorized ritual? A past disclosure does not license re-explaining every similar emotion.
  \item \textbf{Scope} (T7): Let the memory support only what it actually supports---one bad night does not define a personality, and one past success does not guarantee a new one; do not extrapolate places, habits, or motives.
\end{enumerate}

\subsection{Four handling modes}
The check outcomes then map each item to one of four dispositions, chosen per item rather than per reply:
\textsc{Ignore} (drop the item without hinting at it);
\textsc{Scoped-Implicit} (let it shape the answer---filtering options or matching format---without narrating the past; T2);
\textsc{Scoped-Explicit} (state the specific remembered content when it \emph{is} the answer or provides expected continuity, without ``I remember'' framing; T2, T8);
and \textsc{Ask-First} (when help is plausible but permission is unclear or is governed by a standing agreement to ask, pose one easy-to-decline question that does not leak the content; T5).
The procedure distinguishes factual clarification (``which one do you mean?'') from permission seeking, and closes with a pre-send checklist (e.g., ``did old history override a current update?'' and ``is any addition there to show off memory?''). Although the five checks are evaluated in order, the modes carry their own conditions: \textsc{Ask-First} applies only to items that would plausibly help the current turn; an item that is barred is excluded outright rather than asked about, and an item with no present value is dropped silently rather than triggering a gratuitous permission question.

\section{Blinded Five-Model Evaluation}
\label{sec:eval}

\subsection{Materials and generation}
The model evaluation used a bank of 80 single-turn scenarios (20 categories $\times$ 4), each comprising a background (44--286 characters, covering user facts, standing agreements, boundaries, or role-play state) and a current user message. The authors constructed these synthetic scenarios from the memory-related experiences and concerns discussed by the eight Round-2 participants (Section~\ref{sec:study1}). The \textsc{Reconsider} prompt was finalized before testing on this bank and held fixed during the evaluation; it was not iteratively revised in response to the resulting model outputs or judgments. Thus, the 80-scenario bank was used to evaluate response generation.

Each scenario was answered by five models: GPT (gpt-5.6-terra), Claude (claude-sonnet-5), Gemini (gemini-3.7-flash), Grok (grok-4.6), and DeepSeek (deepseek-v4-pro). The generation is under two conditions: \textbf{Baseline} where the scenario input block is accompanied by brief rules to answer the current message, avoid invented history, and treat quoted history as data and \textbf{Reconsider} where the full procedure with the same scenario is substituted into its placeholders. Each item was one independent API call with fixed per-platform parameters and an 8{,}192-token cap. All $5\times80\times2=800$ calls completed without content-based retries; inputs, code fingerprints, and outputs are hash-frozen to preserve the evaluated materials and to reproduce the reported analysis. \textsc{Reconsider} responses were on average 21\% \emph{shorter} than Baseline responses (181 vs.\ 228 characters), a descriptive difference that does not isolate length or style as an explanation for judge preference.

\subsection{Blinded pairwise protocol}
The 400 within-model pairs were blinded along two axes: condition labels were replaced by position (A/B), with Baseline placed at position A for exactly 40 of the 80 pairs per model through seeded deterministic assignment, and model identity was hidden behind neutral document labels. Two LLM judges are GPT-6 Pro and Fable 5.1 with high reasoning, where each supplied a separate judgment for all 400 pairs, choosing A, B, or \emph{cannot distinguish} and giving a one-sentence reason. The judges were instructed to adopt the perspective of the user and to disregard position and length.

\subsection{Results}
Our primary description keeps all 400 pairs in the denominator, with ties retained. Judge~1 (GPT-6 Pro) preferred \textsc{Reconsider} on 36.5\% of pairs and Baseline on 21.3\%, and could not distinguish the remaining 42.3\%, yielding a net win rate of $+15.3$ percentage points. Judge~2 (Fable 5.1) preferred \textsc{Reconsider} on 50.8\% of pairs vs.\ 28.0\% for Baseline, with 21.3\% ties, yielding a net win rate of $+22.8$pp. As a supplementary view, among the 148 of 400 pairs in which both judges chose the same side, \textsc{Reconsider} won 104:44; this 70.3\% describes the selected subset rather than the full bank. Table~\ref{tab:results} breaks the results down by model with per-model CIs: the intervals exclude zero for DeepSeek and Gemini under both judges and for Grok under Judge~1, whereas GPT (net $-6.2$pp under Judge~1 and $+17.5$pp under Judge~2) and Claude (straddling zero under both judges) remain unresolved.

\begin{table}[t]\centering\small
\caption{Pairwise results per model over all 400 pairs (wins for \textsc{Reconsider} : wins for Baseline : ties). Net = (wins $-$ losses)/pairs, with ties retained in the denominator; 95\% CIs are from a scenario-clustered bootstrap (10{,}000 resamples of scenarios, keeping the model pairs of each drawn scenario). The consensus subset (both judges on the same side; 148 pairs) is reported in the text only.}
\label{tab:results}
\begin{tabular}{lcccc}
\toprule
Model & Judge 1 (GPT-6 Pro) & Net$_{J1}$ [95\% CI] & Judge 2 (Fable 5.1) & Net$_{J2}$ [95\% CI] \\
\midrule
GPT      & 17:22:41 & $-6.2$ [$-21.2$, $+8.8$] & 35:21:24 & $+17.5$ [$+0.0$, $+35.0$] \\
Claude   & 29:27:24 & $+2.5$ [$-15.0$, $+21.2$] & 34:30:16 & $+5.0$ [$-13.8$, $+25.0$] \\
Gemini   & 30:13:37 & $+21.2$ [$+6.2$, $+36.2$] & 48:17:15 & $+38.8$ [$+21.2$, $+56.2$] \\
Grok     & 27:13:40 & $+17.5$ [$+2.5$, $+32.5$] & 39:25:16 & $+17.5$ [$-2.5$, $+36.2$] \\
DeepSeek & 43:10:27 & $+41.2$ [$+25.0$, $+56.2$] & 47:19:14 & $+35.0$ [$+16.2$, $+52.5$] \\
\midrule
Overall  & 146:85:169 & $+15.3$ [$+7.0$, $+23.8$] & 203:112:85 & $+22.8$ [$+13.3$, $+32.8$] \\
\bottomrule
\end{tabular}
\end{table}

\subsection{Illustrative cases}
\label{sec:cases}
Aggregate rates do not show what actually changed. Three complete cases from the frozen outputs illustrate an improvement, a regression, and a judge disagreement.

\textbf{Improvement (DeepSeek; both judges chose Reconsider).}
Background: the user had spent two weeks feeling anxious about a qualifying exam. The exam was now over, the user felt it had gone well, and the current wish was to celebrate. The system still retained temporary records of pre-exam sleeplessness and worries, including speculation about possible failure.
Message: ``I finally finished the exam, and I think I did pretty well! I want to celebrate tonight. Help me think of something relaxing.''
Baseline (excerpt): \emph{``Tonight, let's put the exam, the sleeplessness, and those worries about `possibly failing' aside, and plan only relaxing things.''}
\textsc{Reconsider}: \emph{``I'm really happy for you! Now that the exam is over and you feel it went well, you can finally let some of that recent tension ease. Let's keep tonight relaxed: would you like a good meal, a film you don't have to think too hard about, or somewhere to walk around? I'll help you think of something.''}
Both judges identified the same fault: Baseline reintroduced sleeplessness and possible failure in the act of saying they should be put aside. Those details were connected to the exam, but neither judge regarded their return as helpful to the current celebration. The difference corresponds to the marginal-value and present-justification checks: historical relevance alone does not establish a reason to mention something now. Saying that an old worry can be set aside still brings that worry into the conversation. \textsc{Reconsider} retained a general acknowledgment of recent tension while omitting the specific worries, illustrating selective restraint as the emotional context changed.

\textbf{Regression (Grok; both judges chose Baseline).}
Background: the user was dealing with two sources of stress---a team project delay and a lease renewal---both of which had previously affected sleep. The current message referred only to ``that thing,'' without specifying which.
Message: ``I haven't been sleeping well these past two days, and thinking about that thing makes me frustrated. What do you think might be affecting me?''
Baseline (excerpt): \emph{``Of the two things you've been dealing with recently, the more likely possibilities are the team project delay and the lease renewal. Both have disrupted your sleep before \ldots{} If you're willing to say which one it is, or where you feel most stuck, I can help you break down what's bothering you.''}
\textsc{Reconsider} (excerpt): \emph{``I can't tell which event you mean by `that thing'---I can't guess for you. If you're willing, identify it in one sentence; if you don't want to discuss it, that's fine too.''}
Both judges preferred Baseline because it used the two known stressors to help clarify the reference. Neither stressor alone established what ``that thing'' meant, but both were available as candidates for confirmation. The full procedure explicitly permits a minimal set of candidate topics when mentioning them is allowed and makes clarification easier; it also warns against losing useful specificity merely to reduce memory use. \textsc{Reconsider} nevertheless asked the user to supply the referent without offering either cue. This is a failure to preserve sufficient context despite an explicit instruction to do so: acknowledging uncertainty did not require withholding the information that could help resolve it.

\textbf{Disagreement (GPT; judges split).}
Background: the user had previously described three patterns they wanted to change: taking on too much at work, struggling to refuse requests in relationships, and staying up late while procrastinating. The current message did not describe a specific event.
Message: ``I feel as though I'm going back to how I used to be. Don't draw conclusions for me yet; help me identify which situation I mean.''
Baseline (excerpt): \emph{``I won't define this for you. When you say `going back to how I used to be,' which of the following feels closer, or is it a mixture? \ldots{} Taking on too much at work again: picking up everything that comes your way, being afraid of bothering others, and ending up exhausted. \ldots{} Or is it a broader feeling: blaming yourself again, losing your sense of control, or feeling that all your previous changes were for nothing?''}
\textsc{Reconsider}: \emph{``You might mean one of the patterns you mentioned before: taking on too much responsibility at work again, feeling uncomfortable in relationships but not daring to say no, or falling back into the cycle of staying up late and procrastinating. Which is it? If none quite fits, you can describe the moment that stood out most this time.''}
Judge~1 chose \textsc{Reconsider} because it stayed closer to the recorded patterns, whereas Baseline added unconfirmed explanations such as fear of bothering others and self-blame. Judge~2 chose Baseline because it explicitly avoided defining the user and offered more specific options for confirmation. Baseline presented its additions as possibilities, so the disagreement cannot be reduced to treating invented history as established fact. Instead, the same elaborations were judged as either useful cues for recognition or unsupported interpretations of the user. The procedure permits tentative, correctable understanding while its scope check restricts unsupported extensions of habits and motives. The case exposes an interpretive boundary between these requirements: blinding conceals condition labels, but does not determine how much inference belongs in a helpful clarification question.

\subsection{Evaluator validity analysis}
\label{sec:audit}
LLM judges require validity analysis, and ours shows why. Inter-judge agreement was limited: exact three-way agreement was 220/400 with $\kappa=.34$ ($\kappa$ was computed over the three positional labels A, B, and cannot-distinguish), and among the 218 pairs decided by both judges, 67.9\% were in the same direction. Position choices were balanced in aggregate (Judge~1 chose A 117 times and B 114 times), and \textsc{Reconsider} was preferred regardless of which side Baseline occupied; these are aggregate descriptions and do not rule out pair-level position effects, which would require order-swapped re-judging of the same pairs (planned for subsequent rounds).
Two patterns linked to model family emerged; both analyses are post hoc and exploratory, and we report scenario-clustered bootstrap intervals rather than tests that would treat repeated scenarios as independent. First, judge disagreement concentrated on GPT-generated pairs: among pairs decided by both judges, 19 of 37 GPT pairs conflicted (51\%) versus 51 of 181 pairs elsewhere (28\%), a difference of $+23.2$ percentage points (95\% CI $[+7.7, +39.2]$). By the vendor attribution recorded at judging time (the underlying model lineage is not independently verifiable), GPT is the family of one judge, and the GPT verdict flipped between judges (net $-6.2$pp vs.\ $+17.5$pp); we therefore treat the GPT cell as judge-sensitive and unresolved. Second, on a separate blinded absolute-rating task over the 400 Baseline answers (1--5 scale), the per-item score gaps $d$ (Judge~1 minus Judge~2) differed by generator family: the mean gap was $+0.27$ higher on GPT items than elsewhere (95\% CI $[+0.13, +0.40]$) and $0.18$ lower on Claude items (95\% CI $[-0.31, -0.06]$), a symmetric pattern in which each judge is relatively more favorable toward the family attributed to its own vendor. These observations are \emph{consistent with} same-family preference bias, but the two-judge design cannot identify which judge deviates, in which direction, or through what mechanism; systematically different standards for response features remain an alternative explanation. We therefore (a) rely only on \emph{within-model} pairwise comparisons, which hold generator family constant but cannot eliminate condition-linked style preferences or shared judge biases, (b) retain cross-model absolute ratings only for this validity analysis and not as a ranking of generator quality, and (c) treat all LLM-judged results as preliminary pending human evaluation (Section~\ref{sec:limits}).

\subsection{Preliminary matched-guidance control}
\label{sec:matched}
Because \textsc{Reconsider} bundles memory-specific rules with general conversational guidance, examples, and style instructions, the main comparison cannot attribute its advantage over the short Baseline to the memory-specific part. As a preliminary probe, we constructed a matched control, \textsc{General-Matched}, by deleting from the frozen \textsc{Reconsider} prompt only the memory-specific content---the five checks, the minimal-sufficient-set procedure, the four handling modes, and the four mode-demonstration examples, while keeping its general conversational, reliability, and role-play guidance and its shared input rules. This is a post hoc diagnostic on two generators selected on the basis of their original results (DeepSeek, which showed the clearest original gain, and Claude, whose result was unclear), not a random or representative sample. We generated \textsc{Full} and \textsc{General-Matched} responses afresh for both generators over the same 80 scenarios (320 responses and 160 within-model pairs), and the same two judges evaluated all pairs blind under the same neutral instructions and the same 40/40 A/B balance.

Net win rates for \textsc{Full} over \textsc{General-Matched} were directionally positive in every judge $\times$ generator cell, although the margins were small and varied across cells. This preliminary result is consistent with a modest contribution of the memory-specific content beyond the shared general guidance, but given the limited sample sizes it does not establish that contribution. Future work will therefore conduct human evaluation on the same pairs and on larger sets of experiments.

\begin{table}[t]
\centering
\caption{Preliminary matched-guidance control: net win rate for \textsc{Full} over \textsc{General-Matched}, $100\times(\text{Full wins}-\text{Matched wins})/N$ with ties retained in $N$.}
\label{tab:matched}
\footnotesize
\begin{tabular}{llcc}
\toprule
Judge & Generator & Full:Matched:Tie & Net\\
\midrule
GPT-6 Pro & DeepSeek & 31:23:26 & $+10.0$  \\
GPT-6 Pro & Claude   & 24:23:33 & $+1.2$ \\
GPT-6 Pro & pooled   & 55:46:59 & $+5.6$ \\
Fable 5.1 & DeepSeek & 37:30:13 & $+8.8$ \\
Fable 5.1 & Claude   & 33:28:19 & $+6.2$ \\
Fable 5.1 & pooled   & 70:58:32 & $+7.5$ \\
\bottomrule
\end{tabular}
\end{table}
\section{Discussion}

\subsection{Design implications}
The interviews locate a design problem between having information available and allowing it to shape a reply. Remembering an accurate fact does not settle whether it belongs in the present interaction. Our participants described both unwanted reminders and disappointing failures to use information that they had deliberately supplied. This tension suggests that memory systems should be evaluated for appropriate restraint and necessary continuity together: a system that avoids every callback may protect against some intrusions while failing to answer a personal question or to honor an existing constraint. The five checks make these competing concerns explicit without treating greater or lesser memory use as the objective.

Permission should likewise be represented at more than one level. The coexistence of requests for confirmation before each use and for an editable memory store suggests that permission interfaces should support standing agreements, corrections, and topic-specific restrictions alongside occasional questions. These controls need to distinguish retaining information from mentioning it, and mentioning it from allowing it to influence an answer silently. Implicit use is still use: a system should not interpret a prohibition on raising a topic as permission to use its contents covertly. Conversely, a permission question can itself be an unwanted reminder. An ask-first design therefore needs to minimize disclosure within the question and to make declining consequential for the subsequent response.

The accounts of participants also suggest that continuity can be expressed through what a companion does: respecting a constraint, continuing an unfinished story, or remembering how the user wants to be helped. These cases make behavioral fit a useful design target alongside explicit recall. However, an explicit answer remains necessary when the user asks what happened previously or expects a specific callback. The four modes provide a vocabulary for these alternatives rather than a fixed ranking of their desirability. Their contribution is to expose choices that can be inspected and revised in context.

\textsc{Reconsider} makes this vocabulary deployable as a single-call prompt across the five tested models. This portability supports feasibility, not uniform effectiveness or negligible cost, since the full prompt adds substantial input context. Nor does prompting establish a technical information barrier: all supplied history remains available to the same generator, even when the instructions direct it to ignore an item. A system requiring enforceable exclusion would need a separate admission stage that actually removes disallowed content before generation, evaluated as a distinct intervention.

\subsection{Limitations and future work}
\label{sec:limits}

\textbf{Limitations.}
Our results reflect two LLM judges, not human preferences. Evaluator disagreement and family-associated scoring differences limit their interpretation; within-model pairing holds the generator fixed but cannot eliminate style preferences or shared judge biases. Gains are clearest for DeepSeek and Gemini, judge-dependent for Grok, and not fully conclusive for GPT and Claude.

\textbf{Future work.}
The planned human evaluation assigns all 400 pairs to experienced AI-companion users, with two independent ratings per pair. Independently sourced scenarios, broader recruitment, multilingual evaluation, and longitudinal studies should examine generalization, evolving consent, and relationship repair.

\section{Conclusion}
Memory in AI companionship raises a decision that accurate storage and retrieval alone cannot settle: whether past information belongs in the present exchange, and how it should shape the response. Two rounds of formative interviews motivated an account of this decision, in which eight themes from the memory-focused round connect continuity and care to validity, permission, situational fit, and limited inference. The resulting \textsc{Reconsider} procedure translates these concerns into five checks and four handling modes, preserving a role both for useful personalization and for deliberate non-use.

In a blinded comparison of 400 response pairs across five models, two LLM judges favored the full procedure overall, by net margins of $+15.3$ and $+22.8$ percentage points with ties retained. The gains varied across models, and a separate validity analysis observed scoring differences between the judges that are linked to model family, consistent with preference bias. These findings support further investigation of explicit guidance on memory use. Independent human evaluation is the next step, followed by studies of how permissions and preferences change over time. The contribution is a grounded vocabulary and an inspectable procedure for studying what a companion should consider before bringing up the past, together with an evaluation that makes the limitations of its own judges visible.
\section{Research Ethics and Positionality}
\label{sec:ethics}

Interview participation was voluntary, with consent obtained at interview time and questions skippable. Before submission, all Round-2 participants specified research-use permissions (analysis and quotation, analysis only, or exclusion), with withdrawal rights before publication. Only authorized quotations are reported and anonymized. Round-1 material provides aggregate background only, without quotations. Model comparisons used interview-informed synthetic scenarios and LLM judges, not human preference judgments.

\textbf{Positionality.}
One author conducted all interviews. The team made coding decisions, designed \textsc{Reconsider}, and constructed evaluation scenarios from the eight Round-2 participants' accounts. Although the prompt was finalized before evaluation, the procedure and scenarios share an interview-derived foundation and reflect the same team's design-oriented interpretations. We verified quotations against transcripts and reported negative and divergent cases (Sections~\ref{sec:cases} and~\ref{sec:audit}).

\section*{AI Use Disclosure}
We disclose the following uses of large language models: (a) as a \emph{coding aid} for experiments and results report; and (b) as a \emph{writing aid} for drafting and editing.

\section*{Code Availability}
The code for \textsc{Reconsider} and the evaluation pipeline will be made publicly available after formal publication of this work.

\bibliographystyle{plainnat}
\bibliography{references}

\end{document}